\documentclass{aa}

\usepackage{natbib}
\usepackage{txfonts}
\usepackage{graphicx}
\usepackage{ulem}

\def\ciii{C\,{\sc iii]}}
\def\civ{C\,{\sc iv}}
\def\nv{N\,{\sc v}}

\def\feii{Fe\,{\sc ii}}

\newcommand\lsim{\mathrel{\rlap{\lower4pt\hbox{\hskip1pt$\sim$}}
        \raise1pt\hbox{$<$}}}
\newcommand\gsim{\mathrel{\rlap{\lower4pt\hbox{\hskip1pt$\sim$}}
        \raise1pt\hbox{$>$}}}

\usepackage{hyperref}
\hypersetup{
    colorlinks=true,       
    linkcolor=blue,          
    citecolor=blue,        
    filecolor=blue,      
    urlcolor=blue           
}

\begin{document}

\title{Broad-line region configuration of the supermassive binary black hole candidate PG1302-102 in the relativistic Doppler boosting scenario}
\titlerunning{Broad-line region of BBH candidate PG1302-102}
\author{
    Zihao Song \inst{1,2}\and 
    Junqiang Ge \inst{1}\and 
    Youjun Lu \inst{1,2,\dagger}\and
    Changshuo Yan \inst{1,2}\and 
    and Xiang Ji\inst{1,2}
    }
\authorrunning{Song et al.}
\institute{CAS Key Laboratory for Computational Astrophysics, National Astronomical Observatories, Chinese Academy of Sciences, No. 20A Datun Road, Beijing 100101, China; $^\dagger$luyj@nao.cas.cn  \and
School of Astronomy and Space Science, University of Chinese Academy of Sciences, No. 19A Yuquan Road, Beijing, 100049, China}
\abstract{PG1302-102 is thought to be a supermassive binary black hole (BBH) system according to the periodical variations of its optical and UV photometry, which may be interpreted as being due to the relativistic Doppler boosting of the emission mainly from the disk around the secondary black hole (BH) modulated by its orbital motion. In this paper, we investigate several broad emission lines of PG1302-102 using archived UV spectra obtained by IUE, GALEX, and Hubble, to reveal the broad-line region (BLR) emission properties of this BBH system under the Doppler boosting scenario. We find that the broad lines Ly$\alpha$, \nv, \civ, and \ciii\ all show Gaussian profiles, and none of these lines exhibits obvious periodical variation. Adopting a simple model for the BLR, we perform Markov chain Monte Carlo fittings to these broad lines, and find that the BLR must be viewed at an orientation angle of $\sim33^{\circ}$, close to face-on. If the Doppler boosting interpretation is correct, then the BLR is misaligned with the BBH orbital plane by an angle of $\sim51^\circ$, which suggests that the Doppler boosted continuum variation has little effect on the broad-line emission and thus does not lead to periodical line variation. We further discuss the possible implications for such a BLR configuration with respect to the BBH orbital plane.}
\keywords{
line: profiles--galaxies: active--galaxies: quasars: supermassive black holes: individual--PG1302-102
}
\maketitle

\section{Introduction}
\label{sec: Intro}

Supermassive binary black holes (BBHs) are predicted in many galactic centers \citep[e.g.,][]{1980Natur.287..307B, 2020ApJ...897...86C} since larger galaxies are formed by hierarchical mergers of smaller ones \citep[e.g.,][]{2000MNRAS.319..168C, 2002MNRAS.336L..61H, 2015ARA&A..53...51S}, and most galaxies host a supermassive black hole (SMBH) in their centers \citep[e.g.,][]{Magorrian1998, 2013ARA&A..51..511K}. However, it is hard to directly detect BBH systems at subparsec separations because of the limitation in spatial resolutions of current available facilities \citep[e.g.,][]{ 2002MNRAS.331..935Y}.  BBH systems must be selected because of some particular signatures, such as orbital modulated periodical light curves \citep[e.g.,][]{2015Natur.518...74G, 2015MNRAS.453.1562G, 2016MNRAS.463.2145C, 2018MNRAS.476.4617C, 2019ApJS..241...33L}, double-peaked or asymmetric line profiles \citep[e.g.,][]{2011ApJ...738...20T, 2012ApJS..201...23E, 2013ApJ...777...44J, 2014ApJ...789..140L, 2019MNRAS.482.3288G}, deficiency of optical--UV continuum radiation \citep[e.g.,][]{Yan2015, Zheng16}, and broad line polarization \citep{2019A&A...623A..56S}.  More than $100$ BBH candidates have been proposed in the literature according to various observational signatures, and most of them are based on the periodicity in their optical light curves,  \citep[e.g.,][]{2015MNRAS.453.1562G, 2016MNRAS.463.2145C, 2018MNRAS.476.4617C, 2019ApJ...884...36L, 2020arXiv200812329C, 2020arXiv200812317L}.

One of the most intriguing BBH candidates is PG1302-102 at redshift $z=0.2784$, which was monitored for more than $20$ years. Its optical and UV light curves show clear periodical variations with a period of $\sim 1884$\,days in the observer's rest frame, corresponding to $\sim 1474$\,days in the PG1302-102 rest frame \citep[]{2015Natur.518...74G}. The sinusoid shape of its light curves can be well fitted by orbital modulated Doppler boosted emission from the disk around the secondary black hole (BH) in an unequal BBH system \citep{2015Natur.525..351D}. In this scenario, the BBH orbital plane must be viewed at an orientation angle, defined as the angle between the line of sight (LOS) and the normal direction of the BBH orbital plane, $i_{\rm orb} \ge 60^{\circ}$, close to being viewed edge-on, in order to explain the amplitude ($\sim14\%$) of the optical variability of PG1302-102, with a spectral index of $\alpha_{\rm opt} =1.1$ \citep[][]{2015Natur.525..351D, 2020MNRAS.496.1683X}. 

For a BBH system like that proposed for PG 1302-102, if its broad-line region (BLR) is flattened and aligned with the BBH orbital plane and also viewed  at an orientation close to edge-on \citep{2015Natur.525..351D}, then the broad emission lines emitted from it are expected to be double-peaked \citep[e.g.,][]{2014ApJ...789..140L, 2016ApJ...828...68N, 2019ApJ...870...16N}. The  broad-line profiles are also expected to vary periodically because the BLR clouds receive Doppler boosted and attenuated ionizing flux periodically \citep[see][]{Ji+20}. However, whether the BLR is aligned with the BBH orbital plane for BBH systems is not clear. For a circumbinary BLR with a flat disk-like structure, if it is misaligned with the edge-on viewed BBH orbital plane, the asymmetry modulated by the Doppler boosting effect on broad-line profiles may be weaker with increasing offset orientation angles, and the profile could also be Gaussian rather than double-peaked.

To investigate the properties of broad emission lines of PG 1302-102, we collect the archived spectroscopic data from the past 40 years to check  whether the broad-line profiles vary in such a long time interval, and  how the orientation of BLR correlates with the orbital orientation constrained under the Doppler boosting BBH scenario. In Section~\ref{section:Preparation} we describe the spectral analyses and introduce a simplified BLR model for fittings of the broad emission lines. In Section~\ref{section: Result} we analyze the profiles of some broad emission lines, and compare the model fitted BLR orientation angle with that of the BBH system under the Doppler boosting BBH scenario. In Section ~\ref{sec:discussion} we discuss possible implications. In Section~\ref{sec:con} we summarize our main results. 

\section{Data analysis and model fitting}
\label{section:Preparation}

PG 1302-102 has been photometrically monitored for a long time in the optical and UV bands. \citet{2015Natur.518...74G} first discovered the periodical variation of PG1302-102 in the optical band with a period $P_{\rm orb}\sim 1884$\,days by using the Catalina Real-time Transient Survey (CRTS) data, and they suggest this variation may indicate the existence of a BBH in PG1302-102. Since then PG1302-102 has been intensively studied in the literature. \cite{2015Natur.525..351D} proposed that the optical and UV light curves of PG1302-102 may be explained as being due to the Doppler boosting of continuum radiation from an accretion disk associated with the secondary BH rotating around the primary BH in a BBH system, and they obtained $P_{\rm orb} \sim 1996$ days. \cite{2018ApJ...859L..12L} further considered the damped random walk (DRW) process in the analysis of the PG1302-102 light curve by including additional data from All-Sky Automated Survey for Supernovae (ASAS-SN),\footnote{https://asas-sn.osu.edu} and they found $P_{\rm orb} \sim 2026$ \,days.  In this paper we adopt $P_{\rm orb} = 1996$\,days, the value obtained by \citet{2015Natur.525..351D} using the Doppler boosting model to fit the light curves, which  corresponds to $\sim 1561$\,days in the PG1302-102 rest frame.

\subsection{Spectral sample and data analysis}

\begin{figure*}
\begin{center}
\includegraphics[width=1\linewidth]{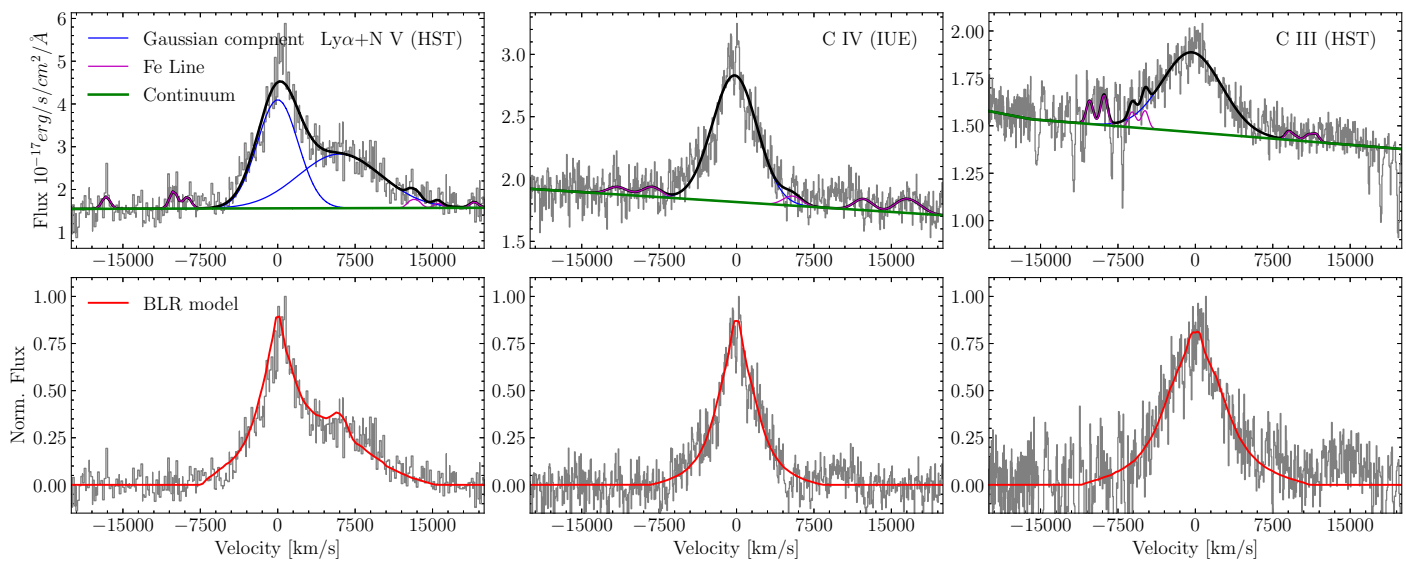}
\caption{Examples of the spectral decomposition (top panels) and the model fitting (bottom panels) to the broad emission lines Ly$\alpha$+\nv\ (left), \civ\ (middle), and \ciii\ (right). In the top panels the observed spectra (gray curves) are fitted by three components, i.e., a power law continuum (green), a Gaussian broad-line component (blue, for Ly$\alpha$+\nv\ we use two broad Gaussian components, with the separation of the two line peaks fixed at $6000 {\rm km s^{-1}}$), and multiple lines from the   \citet{2004ApJ...611...81S} \feii\ templates (magenta). In the bottom panels the gray curves show the broad-line profiles with the continuum and \feii\ lines subtracted. The red lines represent the best BLR model fitting results as introduced in Section~\ref{subsec:model}.}
\label{Fig:fitEg}
\end{center}
\end{figure*}

PG1302-102 has been observed many times in optical and UV bands over the past $40$ years with the International Ultraviolet Explorer  (IUE)\footnote{http://www.vilspa.esa.es/iue/iue.html}, GALaxy Evolution EXplorer (GALEX),\footnote{https://www.nasa.gov/centers/jpl/missions/galex.html} and Hubble Space Telescope (HST)\footnote{https://www.nasa.gov/mission\_pages/hubble}. From the archived spectral data we identify eight Ly$\alpha$+\nv\ broad emission lines (three, two, and three from IUE, GALEX, and HST, respectively), three \ciii\ broad emission lines (two GALEX and one HST), and three \civ\ broad emission lines (two GALEX and one HST), whose peak fluxes are all higher than the $3\sigma$ significance level. These spectra with broad emission lines detected actually include three components: the continuum emitted from the accretion disk, the broad emission line from the BLR, and  \feii\ lines from the BLR. Therefore, we apply a model with three components to fit each observed spectrum, i.e., a power-law for continuum, a Gaussian profile for each broad line, and multiple lines from \feii\ templates \citep{2003ApJS..145...15S, 2004ApJ...611...81S} described by Gaussian profiles with the same velocity dispersion. As done in \cite{2020MNRAS.491.4023S}, the wavelength windows in the rest frame for the fittings to each broad line are $950-1300$\AA, $1180-1280$\AA, $1450-1650$\AA, $1800-2000$\AA, $4250-4650$\AA, and $4800-5200$\AA, and the windows for \feii\ lines are  $950-1195$\AA, $1250-1300$\AA, $1450-1530$\AA, $1570-1650$\AA, $1890-1880$\AA, and $1940-2000$\AA. Three fitting examples are shown in the top row of Figure~\ref{Fig:fitEg}; the panels from left to right correspond to  Ly$\alpha$+\nv, \civ, and \ciii\ broad emission lines.
    
For an active SMBH, the BLR size can be inferred from its optical luminosity by using the empirical relationship between BLR size and optical luminosity as \citep[]{2013ApJ...767..149B}
\begin{equation}
\log (R_{\rm BLR}^L/\textrm{lt-days}) \simeq  1.527+ 0.533 \log \left[
    \frac{\lambda L_{\lambda}(5100\rm \AA)}{10^{44}{\rm erg \ s^{-1}}}\right].  
\label{EQ:BLR}
\end{equation}
The above relationship may also be   valid in the case of an active BBH system if the BLR is far away from the BBH\footnote{As demonstrated below, for the case of PG1302-102 the estimated BLR size is indeed significantly larger than the BBH semimajor axis.} since the broad-line emission is determined by the photon-ionization processes \citep[e.g.,][]{2000ApJ...536..284K}. According to the spectra from the KPNO 2.1 telescope and Gold Spectrograph \citep[]{1992ApJS...80..109B}, we have $\lambda L_{\lambda}(5100\rm \AA) \sim 3.2 \times 10^{45}{\rm erg\,s^{-1}}$, thus $R_{\rm BLR} \sim 213$ light-days (lt-days). The FWHM of the \civ~ is $5089\pm 885\ {\rm km\,s^{-1}}$ according to our fittings to this line, $\lambda L_{\lambda}(1350\rm \AA) \sim 7.8 \times 10^{45}{\rm erg\,s^{-1}}$, then the central mass of the system should be $\sim 10^{9.1}M_{\odot}$, according to the empirical mass estimator based on \civ~ given in \citet{2006ApJ...641..689V}, which is consistent with that estimated by \citet[][]{2015Natur.518...74G}, i.e., $\sim 10^{8.3}-10^{9.4}M_\odot$, calculated by combining the empirical virial mass \citep[][]{2011ApJS..194...45S} and that inferred from the damped random walk (DRW) variability \citep{2010ApJ...721.1014M}.

The separation of the BBH system can be estimated if the total mass ($M_{\bullet\bullet}$) and the orbital period ($P_{\rm orb}$) of the BBH system are known, and it can be given by
\begin{equation}
\label{EQ:a_BBH}
a_{\rm BBH} = 16.4 \times 
\left(\frac{M_{\bullet\bullet}}{10^{9.1}M_{\odot}}\right)^{1/3} 
\left(\frac{(1+z)P_{\rm orb}}{1996\,\rm days}\right)^{2/3} \mbox{lt-days}.
\end{equation}
Adopting the   observed orbital period $P_{\rm obs}=(1+z)P_{\rm orb}=1996$ days obtained from the Doppler boosting based light curve modeling \citep{2015Natur.525..351D} and the estimated mass $\sim 10^{9.1}M_\odot$, we have $a_{\rm BBH} \sim 16.4$\,lt-days. If we adopt the mass range given in \citep{2015Natur.525..351D}, we have $a_{\rm BBH} \sim 8.8-20.6$\,lt-days. If we adopt the Doppler boosting hypothesis to model the periodical light curves of PG1302-102, $M_{\bullet\bullet}$ is required to be $>10^{9.1}M_\odot$ (see also Fig.~\ref{Fig:pdf}), and thus $a_{\rm BBH} \sim 16-21$ lt-days. According to all the above estimates, the ratio of the size of the BLR to the separation of the BBH is $R_{\rm BLR}^{\rm L}/a_{\rm BBH} \sim 10 - 24$, which indicates that the BLR size of PG1302-102 is indeed much larger than the BBH separation, and thus it is reasonable to assume a  circumbinary BLR for PG1302-102 as we do in our  modeling of the BLR region.

\subsection{Simple model for  broad-line fitting}
\label{subsec:model}

The broad emission line profiles are determined by the BLR geometry, kinematics, and structures. The BLR geometry is mainly characterized by the radial and angular distributions of BLR clouds. As described in \citet[]{2014MNRAS.445.3055P, 2014MNRAS.445.3073P}, the radial distribution of BLR clouds can be described as a shifted $\Gamma$-distribution, $r=R_{\rm s} + R_{\rm BLR} F + \beta^{2}R_{\rm BLR}(1-F)g$, where $R_{\rm s}=2GM_{\bullet\bullet}/c^2$ is the Schwarzschild radius, $R_{\rm BLR}$ is the mean value of the shifted $\Gamma$-distribution, $F=R_{\rm in}/R_{\rm BLR}$, $R_{\rm in}$ is the inner radius of BLR, and $\beta$ is the shape parameter of $\Gamma$-distribution $g(\beta^{-2}, 1)$. The angle displacement of the BLR clouds is given by $\theta = \arccos(\cos\theta_{\rm o} + (1-\cos\theta_{\rm o}\times U^{\gamma})$, where $\theta_{\rm o}$ is the opening angle and $U$ is uniformly distributed between 0 and 1 \citep{2014MNRAS.445.3055P, 2014MNRAS.445.3073P, 2018Natur.563..657G}.
By adopting this simple BLR model introduced by \cite{2014MNRAS.445.3055P}, we assume that the BLR clouds rotate around the central BBH system on circular orbits, and constrain the size and structure of the BLR by the following seven model parameters, the same as those in Section~2.3 of \citet{2020MNRAS.491.4023S}: the mean and inner radius of BLR ($R_{\rm BLR}, R_{\rm in}$), radial shape parameter ($\beta$), disk edge illumination parameter ($\gamma$), opening angle ($\theta_{\rm o}$), orientation angle ($i_{\rm BLR}$) defined as the angle between the LOS and the normal direction of the flattened BLR middle plane, and the central SMBH mass ($M_{\bullet\bullet}$).
Here, the opening angle $\theta_{\rm o}$ is defined as half of the angular thickness of the BLR, with $\theta_{\rm o}=\pi/2$ corresponding to a spherical BLR.

To fit the broad emission lines with the BLR model and derive a robust estimation to these seven parameters, we adopt the Markov chain Monte Carlo (MCMC) code ``emcee'' \citep[]{2013PASP..125..306F} to perform the model fitting after a series of prior setups. The prior of each parameter is set as follows. We allow the mean radius $R_{\rm BLR}$ uniformly distributed in the range $[0.2R_{\rm BLR}^{L}, 5R_{\rm BLR}^{L}]$, with the inner radius $R_{\rm in}$ in the range $[0, R_{\rm BLR}]$. The orientation angle $i_{\rm BLR}$ is set to range from $0^{\circ}$ (face-on) to $90^{\circ}$ (edge-on), the open angle $\theta_{\rm o}$ ranges from $0^{\circ}$ (flat disk) to $90^{\circ}$ (spherical BLR). The edge illumination parameter of the BLR disk are in the range $[1, 5]$, and the total binary mass ranges from $[10^{8.3}$ to $10^{9.4} \rm M_{\odot}]$ \citep[]{2015Natur.518...74G}. The bottom panels of Figure~\ref{Fig:fitEg} show three BLR model fitting examples of Ly$\alpha$+\nv (left), \civ\ (middle), and \ciii\ (right) broad emission lines, each with the reduced $\chi_{\nu}^{2}\sim 1$.

\section{Results}
\begin{figure*}
\begin{center}
\includegraphics[width=1\linewidth]{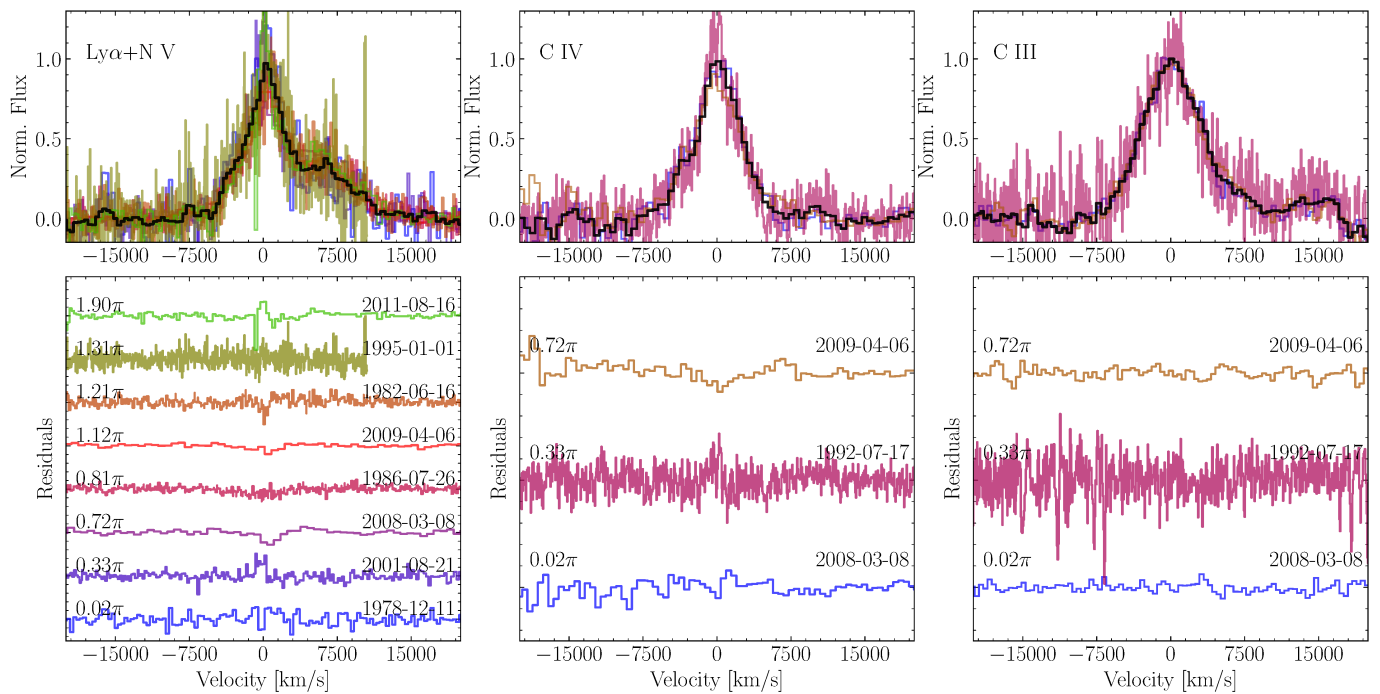}
\caption{Broad emission line profiles of PG1302-102 derived from IUE, GALEX, and HST observations. Columns from left to right show the Ly$\alpha$+\nv, \civ, and \ciii\ broad lines. Top panels show all the observed profiles, normalized to the peak flux of mean broad-line spectrum, with the continuum and \feii\ lines subtracted, while the bottom panels present the corresponding profile residuals of each broad line with the averaged broad emission line subtracted. For each broad-line profile and residuals, the colors from blue to green correspond to spectra observed from $\sim 40$ years ago to recent years, with the exact observation time and phase indicated above each spectrum (top   right and top left, respectively)  in the bottom panels.}
\label{Fig:BLAsym}
\end{center}
\end{figure*}
\label{section: Result}
In order to investigate whether the Doppler boosting effect is reflected by features appearing in broad-line profiles, we first identify how the observed profiles vary at different times, then analyze the MCMC fitted BLR parameters and their correlations with that predicted by Doppler boosting hypothesis of the BBH system, and finally provide constraints on parameters of the binary system possibly existing for PG1302-102.

\subsection{Profile variations of broad emission lines}

According to the periodical light curves in optical and UV bands, PG1302-102 is believed to host a BBH system in the center and should be observed at a close to edge-on orbital orientation, which can lead to Doppler modulated periodical variations of optical and UV luminosities. Supposing that the BLR is aligned with the BBH orbital plane, the emitted broad emission lines should show some double-peaked or flat-topped features, which are characteristic signature of flattened BLR viewed at an orientation close to edge-on. If the secondary SMBH dominates the total luminosity and rotates in relativistic orbital velocity \citep[]{2015Natur.525..351D}, the broad emission line profiles should also be characterized by periodical asymmetry. 

The top panels of Figure~\ref{Fig:BLAsym}   show the mean line profiles obtained by stacking spectra observed at different times over the past 40 year for Ly$\alpha+$\nv\ (left), \civ\ (middle), and \ciii\ (right) broad lines. These broad lines all show Gaussian profiles and do not have any significant evidence for double-peaked, flat-topped, or asymmetric features.\footnote{The signal-to-noise ratios (S/Ns) of the available spectra are not high, which may cause the smearing of such features if they exist. Future higher S/N spectra may further check this.}  
However,  if the BLR is flattened and is aligned with the BBH orbital plane, the broad lines emitted from the BLR should be  double-peaked when viewed at an orientation close to edge-on; if the BLR is spherically distributed, the broad lines should be flat-topped viewed at any orientation. Non-detection of such line features for PG 1302-102 suggests that the BLR is not aligned with the BBH orbital plane.

To further explore the detailed differences, we obtain the spectral residuals for each observed spectrum by subtracting  the corresponding mean spectrum (black profiles in the top row of Fig.~\ref{Fig:BLAsym}), which is obtained as follows. We first normalize the total flux of each broad line, then re-sample the wavelength points of each spectrum to the one with lowest spectral resolution, and finally stack all of them to derive the mean spectrum of each broad emission line. Considering that the archived spectra are observed by different instruments, we hence interpolate the mean spectrum of each broad line to the same wavelength sampling as the observed ones, and then obtain the corresponding spectral residuals, as shown in the bottom row of Figure~\ref{Fig:BLAsym}. For each residual spectrum we label the corresponding phase according to its observation time based on the modeling of optical light curves given by \citet{2015Natur.525..351D}, which would help us to understand if there are any variation trends. To search for possible variation in broad-line profiles, we take the standard deviation of residual spectrum in the ranges   $[-15000,-7500]$ km/s and $[7500,15000]$ km/s as the noise of the residual spectrum. These two wavelength ranges only contain iron lines and show no feature of Ly$\alpha$+\nv\, \civ\ and \ciii\ emissions for the current sample.
Although the spectral residuals fluctuate in the velocity--wavelength space, no spectra have continuing three wavelength points over $3\sigma$ significance level. Perhaps limited by spectral S/N, the spectral residuals at all phases for the three broad emission lines show no significant asymmetric or double-peaked features that can reflect the profile variations that modulated by the Doppler boosting effect.

These results suggest that the Doppler boosting effect, if any, does not influence or modulate the broad-line emission of PG 1302-102. 
This can be explained if the BLR is not aligned with the BBH orbital plane, i.e., the BLR is viewed at an orientation close to face-on, significantly offset from the orbital plane. In this case the ionizing flux received by clouds in the flattened BLR would be Doppler boosted and modulated by the motion of the projected orbital velocity of the secondary BH, i.e., $v_2 \cos \Delta i$ ($\Delta i = i_{\rm orb} - i_{\rm BLR}$), where $\Delta i$ is the offset angle. When $\Delta i$ is large enough, the Doppler boosting effect on the broad emission line profiles would be smaller or even negligible compared to the case with the BLR aligned with the BBH orbital plane, and thus no periodical variation of broad-line profiles would be expected. Further understanding of the BBH system in PG 1302-102 requires more detailed parameters of the BLR geometry and the information of how it is misaligned with the BBH orbital plane.

\subsection{BLR model fitting}
\label{subsec:mResults}

\begin{figure*}
\begin{center}
\includegraphics[width=1\linewidth]{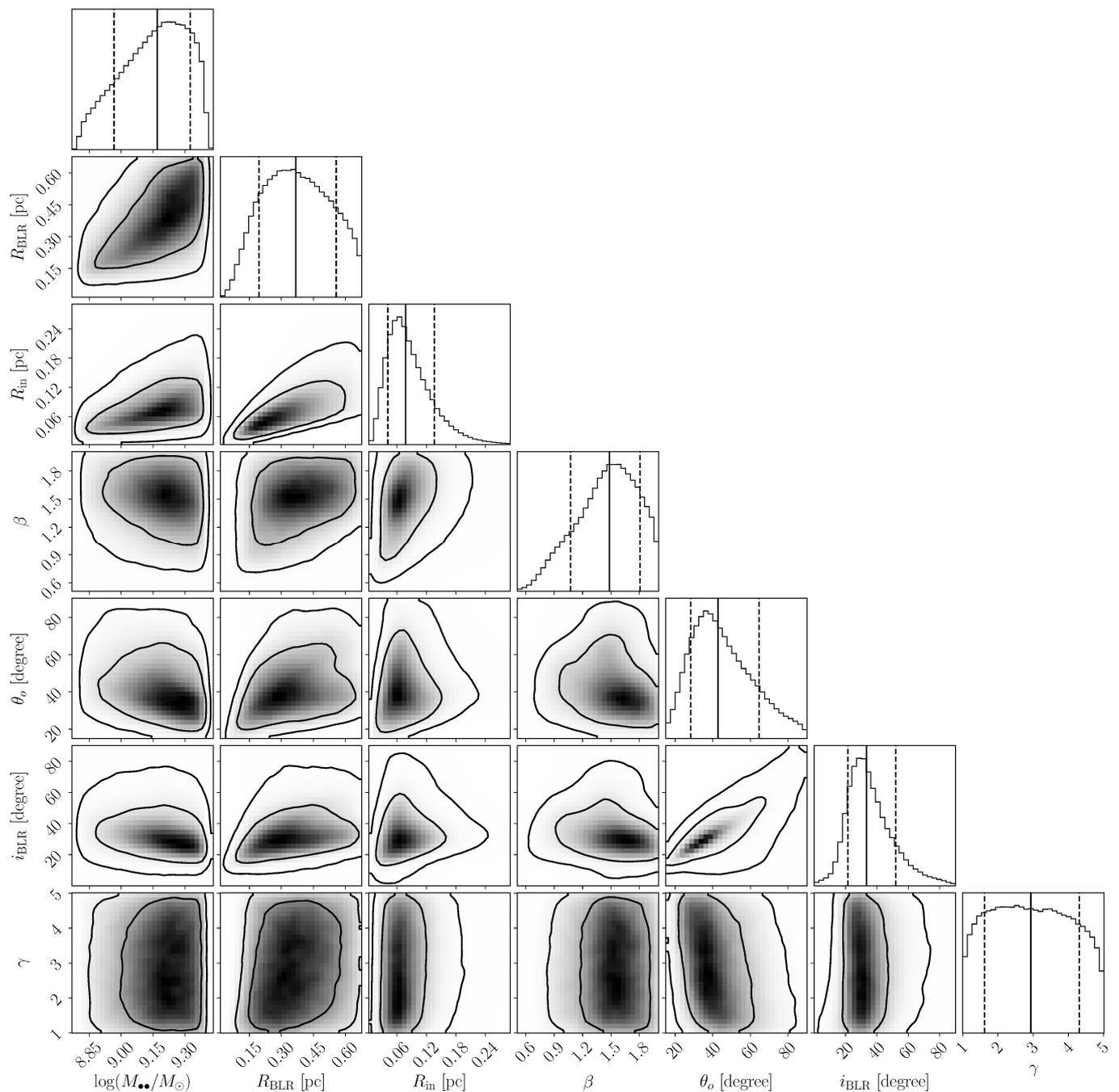}
\caption{
Two-dimensional and one-dimensional probability distributions of the model parameters obtained from the MCMC fittings to the broad emission line profiles with a simple BLR model. The seven model parameters shown here are the total mass $\log(M_{\bullet\bullet})$, BLR size $R_{\rm BLR}$, BLR inner radius $R_{\rm In}$, $\beta$, BLR opening angle $\theta_{\rm o}$, viewing angle $i_{\rm BLR}$, and $\gamma$. In the right side panels, the solid and dashed lines indicate the median, $16$th, and $84$th percentiles of the one-dimensional projected probability distribution function for each model parameter.
}
\label{Fig:pdf}
\end{center}
\end{figure*}
\begin{table*}
\caption{MCMC fitted BLR model parameters for PG1302-102.}
\centering
\begin{tabular}{ccccccc}
\hline \hline
Parameter  & Mean  & Ly$\alpha$ & \nv  & \civ & \ciii & Description \\ \hline 
$\log(M_{\bullet\bullet}/M_{\odot})$  & $9.17^{+0.15}_{-0.20}$  & $9.16^{+0.15}_{-0.19}$ & $9.14^{+0.17}_{-0.20}$ & $9.17^{+0.14}_{-0.23}$ & $9.19_{-0.21}^{+0.15}$ & Total mass of BBHs \\  
$R_{\rm BLR}$ (pc)  & $0.37^{+0.19}_{-0.17}$ & $0.42^{+0.16}_{-0.18}$ & $0.22^{+0.18}_{-0.10}$ & $0.35^{+0.20}_{-0.17}$ & $0.29^{+0.18}_{-0.13}$ & Mean BLR radius \\ 
$R_{\rm in}$ (pc)  & $0.08^{+0.06}_{-0.04}$ & $0.09^{+0.07}_{-0.04}$ & $0.09^{+0.07}_{-0.04}$ & $0.07^{+0.04}_{-0.03}$  & $0.07^{+0.05}_{-0.03}$ & BLR Inner radius \\ 
$\beta$  &  $1.48^{+0.32}_{-0.41}$ & $1.55^{+0.29}_{-0.38}$ & $1.39^{+0.43}_{-0.56}$ & $1.54^{+0.29}_{-0.30}$ & $1.25^{+0.40}_{-0.35}$ & Radial shape parameter \\ 
$\theta_{\rm o}$ ($^{\circ}$)   & $42.8^{+21.8}_{-14.6}$ & $37.8^{+18.1}_{-11.6}$ & $45.2^{+23.6}_{-16.3}$ & $44.5^{+9.88}_{-11.7}$ & $52.6^{+18.8}_{-16.2}$   & Opening angle \\
$i_{\rm BLR}$ ($^{\circ}$)   & $33.3^{+18.6}_{-11.8}$ & $31.4^{+14.4}_{-9.37}$ & $31.5^{+21.4}_{-13.2}$ & $35.0^{+4.04}_{-6.72}$ &$41.2^{+19.1}_{-16.2}$   & orientation angle  \\ 
$\gamma$  & $2.93^{+1.37}_{-1.32}$ & $3.17^{+1.24}_{-1.38}$ & $3.19^{+1.23}_{-1.38}$ & $2.83^{+1.41}_{-1.26}$ & $2.56^{+1.52}_{-1.11}$  & Edge illumination parameter \\ \hline
\end{tabular}
\label{Table:Parameter}
\end{table*}

Similar to our previous work \citep{2020MNRAS.491.4023S}, we adopt the MCMC code ``emcee'' to fit the Ly$\alpha$+\nv, \civ, and \ciii\ broad lines, and obtain the best fit for the model parameters of BLR.

In the BLR model fitting, considering that Ly$\alpha$, \nv, \civ, and \ciii\ lines might be emitted from different BLRs \citep[e.g.,][]{2000ApJ...536..284K}, we hence set the the seven model parameters   used for constructing the BLR as free parameters for the four broad lines. In addition, we also set the shift of line center and flux of the broad line as free parameters. For the joint fitting of Ly$\alpha$ and \nv\ lines, we only bound the shift of line centers for the two lines in the model fitting, with the center of \nv\ red-shifted 6000 km/s from the center of Ly$\alpha$. Each observed line profile can provide a probability distribution function (PDF) for each parameter. For each broad emission line, to improve the robustness of parameter estimation, we stack together all the PDFs (eight for Ly$\alpha$ and \nv, three for \civ, and three for \ciii)  to derive the most robust parameter estimation (see Cols. 3-6 of Table~\ref{Table:Parameter}). All seven of the parameters fitted from the four broad emission lines are consistent with each other in $1\sigma$ confidence level. Therefore, to make the parameter estimation more robust, we also stack all the PDFs derived from eight Ly$\alpha$+\nv, three \civ, and three \ciii\ line fittings together, which produce the final values of the seven model parameters (Col. 2  of Table~\ref{Table:Parameter}).

Figure~\ref{Fig:pdf} shows the stacked PDFs of all the seven parameters in our BLR model, with increasing probabilities labeled from gray  to dark colors, and the two black contours enclose the $68.3\%$ and $95.4\%$ of the population. For each parameter, the two black contours enclose   $68.3\%$ and $95.4\%$  of the PDF. Table~\ref{Table:Parameter} lists the median value with uncertainties represented by the $16$th and $84$th percentiles, which are plotted by vertical lines in the top panels of Figure~\ref{Fig:pdf}. The MCMC fitting of broad emission lines for PG1302-102 predicts a total BH mass of $10^{9.17}M_\odot$ with a typical BLR radius $R_{\rm BLR}\sim 0.37^{+0.19}_{-0.17}$\,pc. The open angle $\theta_{\rm} \sim 43^{+22}_{-15}$ degree and the orientation angle $i_{\rm BLR}\sim 33^{+19}_{-12}$ degree indicate that the BLR is viewed   face-on instead of the edge-on view to the BBH orbital plane \citep[e.g.,][]{2015Natur.525..351D, 2020MNRAS.496.1683X} that is predicted by the Doppler boost hypothesis based on the optical and UV light curves.

\begin{figure*}
\begin{center}
\includegraphics[width=1\linewidth]{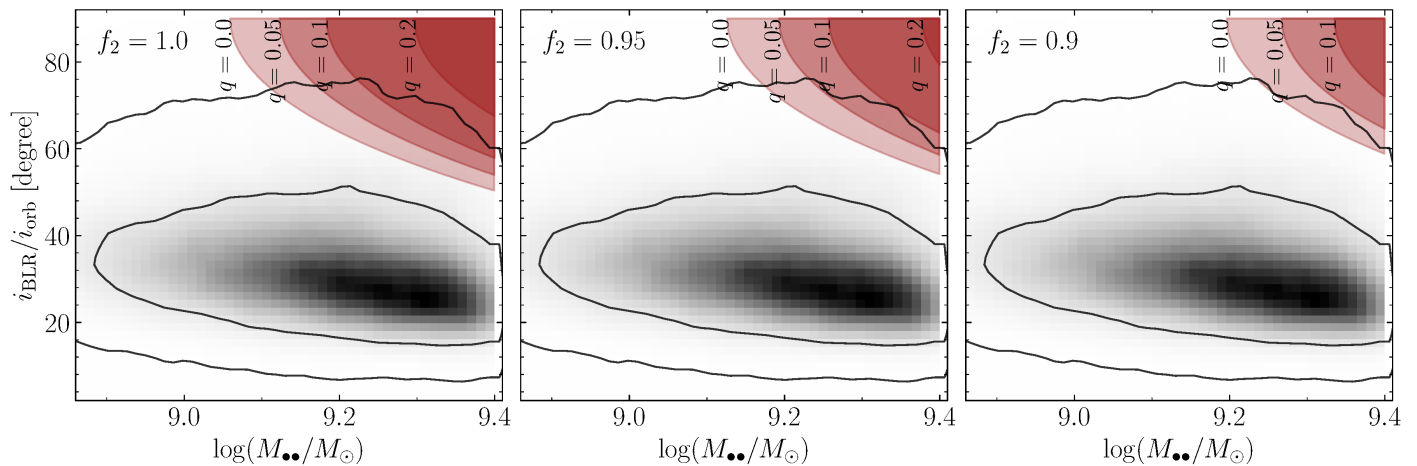}
\caption{
Posterior PDFs for the total mass of BBH  ($M_{\bullet\bullet}$) as a function of $i_{\rm BLR}$ (black) and $i_{\rm orb}$ (red). Panels   show different luminosity fractions of the secondary BH with $f_2=100\%$ (left), $95\%$ (middle), and $90\%$ (right) assumed in the Doppler boosting hypothesis \citep[][]{2015Natur.525..351D}. In each panel, the black shaded region with the $68.3\%$ and $95.4\%$ confidence levels overlapped by black solid lines show the PDF result ($p(i_{\rm BLR},M_{\bullet\bullet})$) obtained from stacking the PDFs of   the Ly$\alpha$+\nv, \civ, and \ciii\ broad emission lines with $S/N>3$. As a comparison, the top right shaded region  correspond to mass ratios $q=0.0$, $0.05$, $0.1$, and $0.2 $ (from pink to dark red).
}
\label{Fig:model}
\end{center}
\end{figure*}
\subsection{Misalignment between the BLR middle plane and the BBH orbital plane}

\begin{figure}
\begin{center}
\includegraphics[width=1\linewidth]{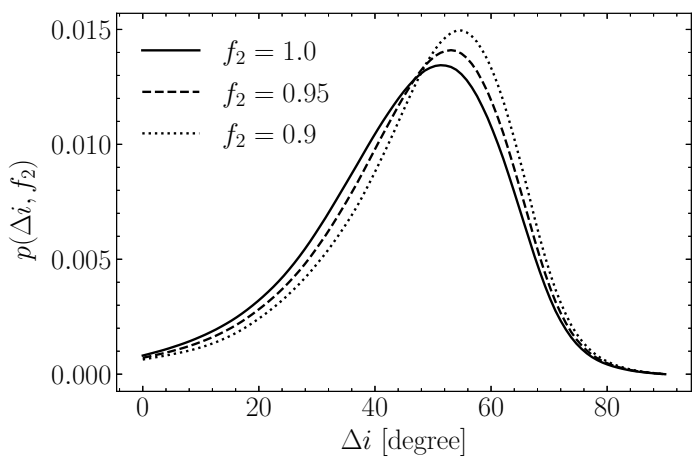}
\caption{Probability distribution of the offset between the orientations of the BLR and BBH ($\Delta i = i_{\rm orb}-i_{\rm BLR}$) assuming a flat distribution of the BBH mass ratio, the total probabilities are normalized to $1$. Solid, dashed, and dotted lines show the cases with a light contribution from the disk around the secondary BH of $100\%$, $95\%$, and $90\%$, respectively.  The estimated peak values of the three probability distributions correspond to $\Delta i= 51^{\circ}$, $53^{\circ}$, and $55^{\circ}$ for $f_2=100\%$, $95\%$, and $90\%$, respectively.
}
\label{Fig:offset}
\end{center}
\end{figure}

Figure~\ref{Fig:model} shows the total mass of BBH $M_{\bullet\bullet}$ as a function of orientation angles for both the BLR and BBH orbital planes, where we can see that the PDF of $i_{\rm BLR}$  (black shaded regions with black contours) are clearly separated  from the $i_{\rm orb}$ distributions (red shaded region), which are the same as that shown in Figure 1 of \citet[]{2015Natur.525..351D}. Panels from left to right show that smaller fractions of the luminosity contribution from the secondary SMBH ($f_2$) actually induce larger differences between $i_{\rm BLR}$ and $i_{\rm orb}$.

To explore the detailed difference between the misaligned $i_{\rm BLR}$ and $i_{\rm orb}$, we calculate the probability distribution of the offset between $p(i_{\rm BLR},M_{\bullet\bullet})$  (black shaded region in Figure~\ref{Fig:model}) and the PDF of BBH orbital plane constrained by the Doppler hypothesis $p(i_{\rm orb}, M_{\bullet\bullet},q, f_2)$ (red shaded region in Figure \ref{Fig:model}) as
\begin{equation}
\begin{aligned}
& p(\Delta i, f_2) =  \\ 
& \iiint p(i_{\rm BLR},M_{\bullet\bullet})p(i_{\rm orb}, M_{\bullet\bullet},q, f_2) d i_{\rm orb} d  \log M_{\bullet\bullet} dq~,
\end{aligned}
\label{EQ:CF}
\end{equation}
where $p(i_{\rm BLR},M_{\bullet\bullet})=p(i_{\rm orb}-\Delta i,M_{\bullet\bullet})$.

\begin{figure}
\begin{center}
\includegraphics[width=1\linewidth]{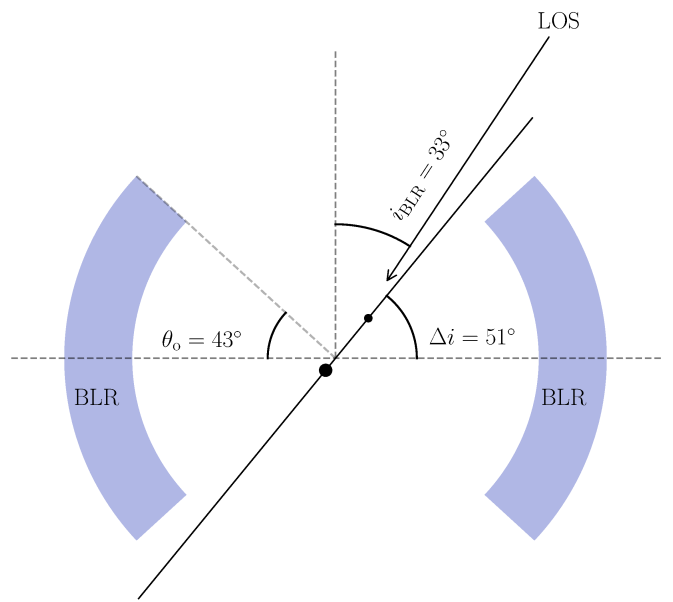}
\caption{Cartoon to illustrate the geometry of a BLR offset from the BBH orbital plane as a possible interpretation for both the broad emission line profiles and periodicity of PG 1302-102 under the Doppler boosting scenario. The inclination angle of circumbinary BLR is $\sim 33^{\circ}$, with an opening angle of $\sim 43^{\circ}$. The BBH orbital plane (solid line) is misaligned from the the middle plane of the BLR with an offset angle of $\sim 51^{\circ}$, which indicates a close to edge-on inclination.
}
\label{Fig:cartoon}
\end{center}
\end{figure}

Given a mass ratio ($q$) of the BBH, we can derive the allowed area $S(q, f_2)$ for the BBH in the $i_{\rm orb}-\log M_{\bullet\bullet}$ plane, as shown by the top right red shaded region in Figure~\ref{Fig:model}. Assuming that the probabilities for each $q$ or $f_2$ are the same, then the probability distribution of $M_{\bullet \bullet}-i_{\rm orb}$ can be defined as

\begin{equation}
\begin{aligned}
\label{EQ:p_def}
p(i_{\rm orb}, M_{\bullet\bullet}, q, f_2) & \propto \frac{p(q)\cdot p(f_2)}{p(S(q,f_2))\cdot S(q,f_2)} \\
& \propto \frac{p(f_2)\cdot \frac{dS(q,f_2)}{dq}}{S(q, f_2)},
\end{aligned}
\end{equation}
where $p(q)$ and $p(f_2)$ are the probability distribution of the mass ratio and the luminosity fraction, and $S(q, f_2)$ is the allowed area for the BBH derived from the relativistic Doppler boosting limitation for the BBH system with given $q$ and $f_2$ as shown in Figure~\ref{Fig:model}. Since the mass ratio is not provided by any other independent method, we assume that $q$ could be any value and that $p(q)$ is uniformly distributed in the range from $0$ to $1$. We also adopt a few fixed values for $f_2$ and $p(f_2)\propto \delta(f_2-C)$ with $C$ a constant. If adopting the mass range $\log (M_{\bullet\bullet}/M_{\odot})\sim [8.3,9.4]$, as given in \citet[]{2015Natur.518...74G}, we  obtain 
\begin{equation}
S(q, f_2)=\int_{8.3}^{9.4} \int_{0}^{1} d\log M_{\bullet\bullet}(q,f_2,i_{\rm orb})d\sin i_{\rm orb}, 
\end{equation}
and the mass of BBHs can be constrained by the optical light curves under the Doppler boosting induced periodical variation scenario as
\begin{eqnarray}
\log\frac{M_{\bullet\bullet}}{M_{\odot}}  & = & 9.1  
+ 3\log\left(\frac{c}{13214\ {\rm km\ s^{-1}}} 
\times \frac{\Delta L_{\rm tot}^{\rm V}}{L_{\rm tot}^{\rm V}} \right. \nonumber \\  
& &  \times \frac{1}{3-\alpha_{\rm opt}} \times\frac{1+q}{1.5} \times \frac{1}{\sin i_{\rm orb}}  \nonumber  \\ 
& & \left. \times\frac{1}{f_{2}-q(1-f_2)}\right) +  \log\left(\frac{(1+z)P_{\rm orb}}{1996\ \rm d}\right) 
\label{EQ:MI}
,\end{eqnarray}
where the power-law index at the optical band $\alpha_{\rm opt}=1.1$. The total luminosity $L_{\rm tot}^{\rm V}$ is contributed by three components, i.e., the mini-disk around the secondary BH and that around the primary BH, and a circumbinary disk (CBD), hence it can be written as
$L_{\rm tot}^{\rm V} = L_{1}^{\rm V}+L_{2}^{\rm V}+L_{\rm CBD}^{\rm V}$. Considering that the primary BH is Doppler modulated by a line of sight velocity $v_{1}\sin i_{\rm orb} = -qv_{2}\sin i_{\rm orb}$, and if we assume that the contribution of CBD is constant over time (see Methods in \citealt[]{2015Natur.525..351D}), then we   have 
\begin{eqnarray}
\frac{\Delta L_{\rm tot}^{\rm V}}{L_{\rm tot}^{\rm V}} 
&=& \frac{\Delta L_{1}^{\rm V}+\Delta L_{2}^{\rm V}}{L_{\rm tot}^{\rm V}} \nonumber \\
&=& (3-\alpha_{\rm opt})\frac{v_{2}\sin i_{\rm orb} }{c}\left[f_{2}-q(1-f_2)\right].
\label{EQ:Lumi}
\end{eqnarray}
Here the rotation velocity of the secondary BH $v_2$ can be obtained by assuming a circular orbit for the BBH system, as done in \citet[]{2015Natur.525..351D}:
\begin{equation}
v_{2}  =  13214 \left(\frac{1.5}{1+q}\right) 
    \left(\frac{M_{\bullet\bullet}}{10^{9.1}M_{\odot}}\right)^{\frac{1}{3}}
    \left(\frac{(1+z)P_{\rm orb}}{1996\, \rm day}\right)^{-\frac{1}{3}} {\rm km s^{-1}}
.\end{equation}
Figure~\ref{Fig:offset} shows the probability distribution of the orientation angle offset $\Delta i$ in the case of $f_2=1.0$, $0.95$, and $0.9$. For the luminosity fraction of the secondary BH $f_2$ decreasing from $1.0$ to $0.9$, the offset between $i_{\rm orb}$ and $i_{\rm BLR}$ increases from $\Delta i=51^{\circ}$ to $55^{\circ}$, which indicates that relativistic Doppler boosting effect is weakened compared to the aligned case.
Figure~\ref{Fig:cartoon} illustrates the geometry of the BLR and BBH system, which shows that the BLR, with an opening angle of $\theta_o=43^\circ$ and viewed by an inclination angle of $i_{\rm BLR} = 33^\circ$, is misaligned with the BBH orbital plane by $\Delta i = 51^\circ$.

The misalignment between the BBH orbital plane and the circumbinary BLR may explain the apparent contradiction between the periodical variation of optical and UV light curves due to the orbital motion of close to edge-on viewed BBH and the Gaussian  profiles of  broad emission lines from flattened BLR viewed at an orientation close to face-on (see  bottom  panels of Fig.~\ref{Fig:BLAsym}). Although the optical and UV continua from the BBHs system vary significantly and periodically due to the relativistic Doppler boosting, the ionizing fluxes received by the most of the BLR clouds do not get significantly boosted by the relative motion between the continuum source and the clouds if the BLR middle plane is offset from the BBH orbital plane. Therefore, the Doppler boosting effect on the broad-line emission line profiles is insignificant, and the line variation is much weaker compared with the continuum evolution. It requires  much higher spectral S/N than the current values to identify the Doppler boosting modulated broad-line profile variations.

\section{Discussions}
\label{sec:discussion}

The offset of the BLR middle plane from the BBH orbital plane may encode some information about the merger configuration of the two progenitor galaxies and their nuclear activities. In principle, the two BH components approach each other at a random orientation, and the merger configuration of the two systems that spiral toward  each other could be one of the following: 1) two comparable BHs, each having a disk and a BLR\footnote{Here and throughout the middle plane of each progenitor BLR is assumed to be the same as the disk plane.}; 2) two comparable BHs, one having a disk and a BLR, but the other not; 3) a small BH and a big BH, both having a disk and a BLR; 4) only the small secondary BH having a disk and a BLR. For each of these cases, when the two BHs become close enough, the system may form a (warped or misaligned) circumbinary disk with the central region cultivated to be a hole or gap by the BBH, within the hole or gap each BH component may still have a small accretion disk surrounding it, and the size of the hole or gap in the case that the two systems merge in a direction inclined to their disk plane is smaller than in their disk plane \citep[e.g.,][]{2013MNRAS.436.2997D, 2014ApJ...783..134F, 2020ApJ...889..114M}. In the cases where the two BHs have significantly different masses, the disk around the secondary BH may dominate the total luminosity of the quasi-stellar objects (QSOs) \citep[see, e.g., ][]{2015Natur.518...74G, Yan2015, 2016MNRAS.463.2145C}. Therefore, the disk orientations would also encode some information about the mergers, but this is beyond the scope of this paper.

For the above four cases, the BLR structure evolves with the orbital decay of the BBH, and the resulting BLR structures from different cases when the BBH system becomes PG1302-102-like may be significantly different. For the first case, the two original BLRs may mix with each other to form a circumbinary BLR, but the initial kinematic structures may be still maintained and thus the profiles of broad emission lines from such a system may be complicated (e.g., with multiple peaks). For the second and fourth cases the resulting system would have a circumbinary BLR with the middle plane offset from the BBH orbital plane as the two systems normally merge in a direction not on the disk plane, and these two planes can be the same only if the merger direction is roughly on the BBH orbital plane. There may be some differences between the BLR structure resulting from these three cases, but it needs carefully designed simulations to reveal it, which is beyond the scope of this paper. In the third case, the middle plane of the resulting circumbinary BLR would be mainly contributed by the components from the original BLR of the big BH and it may   also be normally offset from the BBH orbital plane; the contribution from the original BLR components of the small BH are expected to have some small effects on the broad-line profiles and its significance may depend on the relative number of BLR clouds in the two original BLRs. With these cases in mind, we can use the geometric configurations of the BLR in the BBH systems to (statistically) constrain the merger configuration of their progenitor systems.   

The significant offset between the BLR middle plane and the BBH orbital plane found for PG1302-102 in this paper, assuming the Doppler boosting scenario, indicates that the configuration of the two progenitor BH systems is unlikely to be the first case, but it can be the other three cases with two BHs inspiraling in a direction highly inclined to the disk plane of the big BH in the second and third cases, and to the disk plane of the small BH in the third case. Future observations and their comparison with detailed simulation results on the BLR configuration would help to test the Doppler boosting hypothesis for optical periodicity of some QSOs and extract important information on the merger configuration of the BBH systems.

The derived offset between the BLR middle plane and the BBH orbital plane is based on the Doppler boosting assumption, which can produce persistent periodicity in the light curves. However, the optical--UV light curves of PG 1302-102 may be also alternatively explained as being due to (1) a small chance out of a large sample of Quasars with variability due to the DRW processes \citep[e.g.,][]{2016MNRAS.461.3145V}, (2) the combination of the DRW and periodic variations or the combination of the variation due to a broken power-law power spectrum and the periodical variation \citep{2018ApJ...859L..12L, 2009ApJ...698..895K}, or (3) the cold-spot in the accretion disk around the more massive black hole of the supermassive BBH system \citep{2019ApJ...871...32K}. For these alternative interpretations, the response of the broad emission lines to the continuum variation and the geometry of the BLR   deserves further investigations, which  is beyond the scope of the present paper.
\section{Conclusions}
\label{sec:con}

The most intriguing BBH candidate among those suggested by the optical  periodicity 
is PG 1302-102, which can be  fit well by a model of compact unequal-mass binary system under the relativistic Doppler boosting hypothesis \citep{2015Natur.525..351D}. In this paper we investigated the properties of broad emission lines of PG 1302-102 using the archive spectroscopic data, including the shapes and variations of Ly$\alpha+$\nv, \civ, and \ciii\ lines, and the corresponding parameters of the BLR in order to check whether it is compatible with the Doppler boosting hypothesis. The available observations show that the Ly$\alpha$, \civ, and \ciii\ all have Gaussian profiles, and the multiple observations of these lines do not suggest significant variations of their profiles, even though the variations, if any, may be smeared out due to the low S/Ns and/or spectral resolutions of these observations. Using a simple BLR model we find that the Gaussian profiles of these broad emission lines suggest that the BLR is flattened with an opening angle of $\sim 43^\circ$ and viewed at an inclination angle of $\sim 33^\circ$, close to face-on, and thus not aligned with the BBH orbital plane, which is viewed close to edge-on, constrained by the optical and UV periodical variations. If the middle plane of the BLR is offset from the BBH orbital plane by an angle of $\sim 51^\circ-55^\circ$, the periodical optical--UV variations and the broad emission line profiles of PG 1302-102 can be self-consistently explained under the Doppler boosting scenario.
In this case the misalignment between the BLR middle plane and the BBH orbital plane leads to a highly weakened Doppler boosting effect on the ionizing flux received by the BLR clouds, thus the variations of  emission lines are also  much weaker than the continuum received by the observer. This may be taken as a signature to falsify the Doppler boosting scenario for interpreting the periodical optical--UV variations of PG 1302-102 by using multiple high S/Ns and high resolution spectroscopic observations in the future.

\section*{Acknowledgements}
This work is supported by the National Key Program for Science and Technology Research and Development (Grant No. 2016YFA0400704), the National Natural Science Foundation of China (Grant number 11690024, 11873056 and 11903046), the Strategic Priority Program of the Chinese Academy of Sciences (Grant No. XDB 23040100), and by the Beijing Natural Science Foundation (No. 1204038).

%

%

\end{document}